\documentstyle[12pt,preprint]{aastex}

\begin{document}

\title{The origin of steep vertical stellar distribution in the
Galactic disk\\}
{}

\author{Arunima Banerjee \&  Chanda J. Jog}
\affil{Department of Physics, Indian Institute of Science, Bangalore 560012,
  India, \\e-mail: arunima$\_$banerjee@physics.iisc.ernet.in, cjjog@physics.iisc.ernet.in}

\begin{abstract}
Over the past two decades observations 
have revealed that the vertical density distribution of stars in galaxies 
near the mid-plane 
is substantially steeper than the sech$^2$ function that is expected
from an isothermal approximation.
 However, the physical origin for this has not been explained so far. Here 
 we show that such steep profiles result naturally even within 
 the isothermal regime, on taking into account the 
gravitational force due to the gas.
Due to its low velocity dispersion the gas is concentrated closer to the 
galactic mid-plane than the stars, and hence 
it strongly affects the vertical stellar distribution
even though its contribution to the total surface density is small.
We apply a three-component galactic disk model consisting of gravitationally 
coupled stars and the HI and H$_2$ gas, embedded in the dark matter halo, 
and calculate the vertical density distribution of stars for the Galaxy. 
The resulting vertical  density distribution of stars is shown to be steeper 
than the sech$^2$ function, and lies between the
sech and an exponential function, in good agreement with
observations of galaxies. 
We also show that a multi-component stellar disk consisting of coupled dwarfs and two populations of giants does not explain the observed steep stellar profiles.
\end{abstract}

\section{Introduction}
It is well-known that a gravitating, isothermal stellar disc in a galaxy
can be represented by a self-consistent vertical distribution that 
obeys a sech$^2$ profile 
as was shown theoretically (Spitzer 1942); and confirmed by observations 
(van der Kruit \& Searle 1981 (a,b)). However, recent studies have
shown that the observed vertical
distribution in galaxies is steeper, and is  well-approximated by an
exponential or a sech function,
especially close to the galactic mid-plane, as shown for the Galaxy 
(Gilmore \& Reid 1983;
Pritchet 1983; Kent, Dame \& Fazio 1991; Gould, Bahcall \& Flynn 1996) 
and also for external galaxies 
(Wainscoat et al. 1989; Aoki et al. 1991; Barnaby \& Thronson 1992; 
van Dokkum et al. 1994; Rice et al. 1996). These 
later, near-infrared observations, which are free from the errors due to dust 
extinction, allow one to probe regions closer to the mid-plane. These show an 
excess over the isothermal case near the mid-plane,
 which is better-fitted by an exponential or a sech profile.

 The sech$^{2}$ profile results under the
isothermal assumption for the equation of state so that the
pressure $p$ can be written as density $\rho$ times the square of the sound
velocity or the random velocity dispersion, and the additional assumption 
that the velocity dispersion is independent of vertical distance 
(Spitzer 1942, Bahcall 1984). An exponential distribution as observed in some 
cases was argued to be unphysical when extended all the way to the mid-plane, and a
fit intermediate between an exponential and an isothermal was
suggested in terms a family of curves denoted by a parameter $n$
(van der Kruit 1988). 
Subsequent observational papers analyze the data and give their 
results in terms of this parameter $n$ - see e.g.,
 de Grijis \& van der Kruit (1996); de Grijis,
Peletiers \& van der Kruit (1997). However,  the set of models 
proposed by van der Kruit (1988) is somewhat ad-hoc since it is not obtained 
as a solution to the basic equations relevant in this context.

The physical origin of the steeper-than-isothermal vertical profile 
has not been fully explained so far. An exponential profile is shown
to result  if 
the gas settles in a protogalaxy into isothermal equilibrium and if the star 
fromation rate is equal to the cooling rate of the gas (Burkert \& Yoshii 1996). 
However, this
model requires a specific set of initial conditions, and does
not take account of the secular heating (e.g., Binney \&
Tremaine 1987) that stars undergo,
and also it cannot explain the sech behaviour seen in many galaxies.

In this paper, we show that the steep vertical profile occurs naturally 
on taking account of the strong constraining effect of the gravitational force due to gas.
Because of its lower velocity disperson the gas forms a thin layer and thus lies
closer to the galactic mid-plane than the stars, hence despite its 
small mass-fraction the inclusion of gas is shown to significantly affect the 
stellar vertical distribution.
We apply a three-component model of a galactic disk
consisting of gravitationally coupled stars, HI and H$_2$ gas,
in the field of a dark matter halo, developed earlier by 
Narayan \& Jog (2002 b). The resulting thickness was defined as the HWHM of 
the vertical distribution for each component and it showed a good
agreement with obsevations for the inner Galaxy (Narayan \& Jog
2002 b), and also the moderate stellar flaring predicted agreed
well with observations of external galaxies (Narayan \& Jog 2002
a). In the current paper, we concentrate on the density distribution close
to the mid-plane. There is no discrepancy in these two approaches 
since at higher vertical distances, the various functional forms have the same (sech$^2$) 
behaviour. We treat the stellar disk to be a single component for simplicity,
this assumption is justified later.

The results obtained in this paper show that the
inclusion of gas gravity can account for the observed
steepness in the stellar profiles closer to the mid-plane. Also,
we predict the surface brightness distribution in the K-band at
different galactocentric radii within the Galaxy, considering it as an 
edge-on system as it would appear to an external observer.

In Section 2, we discuss the model used and the
method for solving the equations, and give the parameters used. 
Section 3 contains the results obtained, 
Sections 4 and 5 give the discussion and the conclusion respectively.

\section{Formulation of Equations and Solutions:}
We have used the three-component, galactic
disk model of Narayan \& Jog (2002 b) to obtain the vertical density
distribution of stars in the Galaxy, as discussed in Section 1.
In this model, the response of each component to
the joint potential determines the net
vertical density distribution in a self-consistent manner.
The atomic and
molecular hydrogen gas are distributed in two thin disks, embedded within
 the stellar disk, and are also axisymmetric and coplanar with
it. We use the galactic cylindrical co-ordinates ($R, \phi, z$).
 The equation of
hydrostatic equilibrium in the vertical direction is given by (Rohlfs 1977):
$$
 \frac{<(v_z)_i^2>}{\rho_i} \frac{d\rho_i}{dz} = (K_z)_s + (K_z)_{HI} + (K_z)_{H_2} + (K_z)_{DM} \eqno(1) $$

\noindent where $i$ = 1,2,3, and DM denotes these quantities for 
stars, HI and H$_2$ gas, and the dark matter halo respectively.
Here $K_z = - \partial \psi / \partial z $ is the force per unit mass
along the $z$, and $\psi_i$ is the corresponding potential, and 
$<{(v_z)_i^2}>^{1/2}$ is the random velocity dispersion along $z$.
We have assumed the equation of state to be isothermal and the dispersion to be constant with $z$.

Under the assumption of a thin disk, the joint Poisson equation  
simplifies to:

$$\frac{d^2\psi_s}{dz^2} + \frac{d^2\psi_{HI}}{dz^2} +
\frac{d^2\psi_{H_2}}{dz^2} = 4\pi G(\rho_s+\rho_{HI}+\rho_{H_2})
\eqno(2) $$

\noindent On combining equations (1) \& (2), the density distribution 
of a component $i$ at any radius is given by

$$\frac{d^2\rho_i}{dz^2} = \frac{\rho_i}{<(v_z)^2_i>} \times [ -4\pi
G(\rho_s+\rho_{HI}+\rho_{H_2})+ \frac{ d(K_z)_{DM}}{dz} ] $$
$$ \quad \quad \quad \quad \quad +
\frac{1}{\rho_i} (\frac{d\rho_i}{dz})^2    \eqno(3) $$

From the above equation, it is evident that though there is a
common gravitational potential, the response of each component
will be different due to the difference in their random velocity
dispersions.

\subsection {Solution of the equations}
The above equation represents a set of three coupled, second order
ordinary differential equations which is solved numerically by the
Fourth order Runge-Kutta method in an iterative fashion (Narayan
\& Jog 2002 b) with the following two boundary conditions at the
mid-plane i.e z=0 for each component:

$$ \rho_i = (\rho_0)_i  \qquad \frac{d\rho_i}{dz} = 0  \eqno(4) $$

\noindent However, the modified mid-plane density $(\rho_0)_i $ for each
component is not known a priori. Instead the net surface 
density $\Sigma_i(R)$, 
given by twice the area under curve of
$\rho_i(z)$ versus z, is used as the second boundary condition,
since this is known observationally.
Hence the required value of $(\rho_i)_0$
can be determined by trial and error method, which eventually
fixes the $\rho_i(z)$ distribution.

\subsection {Parameters used}
We require surface density and the stellar velocity dispersion for
each component to solve the coupled set of equations
 at each radius, these parameters have been taken as in Narayan and Jog 
(2002 b). The observed values were used for the gas surface 
densities (Scoville \& Sanders 1987), see Table 1.
 The stellar surface density, and the density profile of the dark matter halo, 
were taken from the standard mass model for
the Galaxy by Mera et al (1998): here
the stellar disk has an exponential distribution with a disk scalelength
of 3.2 kpc and the local stellar density of 45 M$_{\odot}$ pc$^{-2}$, which gives the
central stellar surface density of 641 M$_{\odot}$ pc$^{-2}$, and the dark matter 
halo is taken as an isothermal sphere with a core radius of 5 kpc and the rotation
 velocity of 220 km s $^{-1}$. Table 1 also lists the values of the ratio
 $\Sigma_{HI + H_2} /\Sigma_s$, the total gas to stellar surface mass density. 

The observed gas velocity dispersion values are taken, with 8 km s $^{-1}$ for HI
(Spitzer 1978) and 5 km s $^{-1}$ for H$_2$ gas  (Stark 1984, Clemens 1985). 
 The observed planar stellar velocity dispersion
values from Lewis \& Freeman (1989) were used to obtain the vertical dispersion 
values assuming the same ratio, namely 1/2, of the velocity dispersions as seen in 
the solar neighborhood (Binney \& Merrifield 1998). This gives an exponential 
velocity distribution with a scalelength of 8.7 kpc and the local value of 
18 km s $^{-1}$ (Lewis \& Freeman 1989, Narayan \& Jog 2002 b).

\section{Results}

\subsection {Effect of gas on stellar vertical distribution}

We first illustrate the dynamical effect of the inclusion of the low-dispersion component, 
namely the gas, on the vertical distribution of stars. 
First compare the gravitational force due to stars and gas taken separately.
The strength of the force per unit mass $\vline K_z \vline$ 
for an isothermal single component is obtained by solving the force equation and 
the Poisson equation (eqs. 1-2) together. These can be solved analytically and give 
(Spitzer 1944, Rohlfs 1977):

$$  \vline K_z \vline \: = \: - 2 \frac{<v^2>}{z_0} \: tanh z / z_0   \eqno (5) $$

\noindent where $<v^2>$ is the mean square of the velocity. The vertical 
density  distribution  is given by

$$ \rho (z) \: = \: \rho_0 \: sech^2 (z / z_0)  \eqno (6) $$

\noindent where the constant z$_0$ is given in terms of the mid-plane density, $\rho_0$, as:

$$ z_0 \: = \: \large[\frac{<v^2>}{2 \pi G \rho_0}\large]^{1/2} \eqno (7) $$

We treat the stars-alone case first and use the total observed stellar surface density $\Sigma_s$ at a given radius as a boundary condition, and
integrating eq. (6), we constrain  both $\rho_0$ 
and z$_0$ simultaneously. The
procedure is then repeated for the gas-alone HI and H$_2$ components taken separately. 
The resulting values for the $\vline K_z \vline$ 
the strength of the force per unit mass  due to stars-alone 
and that due to HI and H$_2$ alone (drawn on a log scale) vs. z are shown in Fig. 1, as calculated at R=6 kpc.
It is clear that upto $\vline$ z $\vline < 150 $ pc, the force due to gas is significant 
$\sim 30 \%$ of that due to stars. This is the spatial region over which we 
have fitted the n(z) profiles to obtain the parameter n (Section 3.2).
 Thus we expect to see steepening of the density profile of stars on taking account of 
the gas gravity.

A similar plot for the coupled case cannot be given since
the three disk components (stars, HI and H$_2$) are coupled, hence
the individual terms on the r.h.s. of equation (1) are not known.
 However, a gradient of $\vline K_z \vline$ with z can be given in 
terms of the mass density, $\rho$, for each component using the Poisson equation.
 The resulting values of gas densities 
for the three coupled components as obtained from our model at R=6 kpc are plotted 
in Figure 2. This is another indicator of the importance of gas in the 
dynamics close to the mid-plane. The mid-plane density of the gas and the stars
are comparable, hence the gas 
gravity affects the stellar dynamics significantly.

The dynamical effect of gas gravity on the net vertical density 
distribution of stars is clearly shown in Figure 3, where we plot and compare the stellar density distribution vs. 
z for the stars-alone case and also as obtained in the coupled system.  First, the central or the mid-plane peak in density is increased, and 
 the shape of the central profile becomes sharper which is studied 
quantitatively in terms of the parameter "n" in the rest of the paper.   
Further, the scale-height (HWHM) of the stellar distribution is lower in 
the coupled system due to the inclusion of gas gravity - this 
 aspect was studied in detail in our earlier work (Narayan \& Jog 
2002 b). Thus the inclusion of gas affects the central 
peak density of the stellar distribution, and the shape of the profile, 
and also it decreases the stellar vertical scaleheight.

Thus we have shown that due to its lower velocity dispersion, the gas forms a thinner 
layer and is concentrated closer to the galactic mid-plane than the stars. Hence, 
despite the fact that the gas contains a small fraction of the total disk surface density,
it has a significant effect on
the vertical density distribution of the main mass component, namely the stars.

     A study highlighting a similar constraining effect of a molecular 
cloud complex was done by Jog \& Narayan (2001). The cloud complex is massive
($\sim 10^7 M_{\odot}$)  and extended ($\sim 200$ pc in size), with an average
density much higher than the mid-plane stellar density,
 hence in that case the $\vline K_z \vline $ due
to the complex dominates over the $\vline K_z \vline $ due to stars-alone 
case (Fig. 1 in that paper), and the resulting changes in the vertical 
density distribution of stars and the HWHM (Figs. 3 and 4 respectively in that paper)
are much stronger. Figures 1 and 3 in that paper are the analogs of 
Figures 1 and 3 given above.

\subsection {Model Stellar Vertical Profiles}

The self-consistent vertical density distribution of stars was determined 
for the radial range of 2-12 kpc in the Galaxy, 
following the method discussed in Section 2.1.
In order to compare our results with the observational
papers in the literature which have fitted their data to the van der Kruit (1988) curves, we also fit our
numerical solution for $\rho(z)$, the stellar mass density,
 to the family of curves suggested by van der Kruit (1988):

$$\rho (z) = 2^{-2/n} \: \rho_e \:  sech^{2/n}(nz/2z_e) \eqno (8) $$

\noindent which is characterized by
three  parameters $\rho_e$, $z_e$ and n. Here $\rho_e$ is the extrapolated 
outer mass density away from the galaxy plane which is the same for all 
values of $n$ at large z , and $z_e$ is the vertical scale parameter. 
For a given $\rho_e$, the effective scale-heights and the
amplitudes at z=0 are obtained as a function of the parameter $n$.

In the limiting cases of n=1 and n=$\infty$, eq. (8) reduces to the
usual isothermal (or the sech$^{2}$ function), and an exponential function
respectively, while n=2 denotes an intermediate case of the sech profile.

In Figures 4 and 5, we compare the vertical stellar-density
profiles for n=$\infty$(exp), n=2(sech) and n=1(sech$^{2}$) cases with
our model curves obtained 
at R=6 kpc and 8.5 kpc respectively. The values of
the parameters $n$ and $z_{e}$ in these cases are the best-fitting values
obtained from fitting our model curves to within
 $\vline z \vline \leq$ 150 pc of the galactic mid-plane.
The model curve in Figs. 4 and 5 is best-fitted by n = 6.9 and n = 4.3 respectively,
thus it lies between the sech and an exponential profile. In the region of the 
molecular ring, which includes R= 6 kpc, the influence of gas is higher 
and hence the resulting vertical profile is steeper than in the
solar neighbourhood region of 8.5 kpc. This
underlines the crucial effect of gas in causing a steepening of the vertical stellar density profile in the galactic disk.
The best-fit values for $z_e$ are 368 pc and 362 pc respectively. 
From these figures, it is clear that the best-fitting value of $n$ 
 depends on the range of z used, and at large enough $z$-range all curves go over to the usual
sech$^2$ profile.

The effect of gas gravity is further
illustrated by plotting the best-fitting $2/n$ values versus radius 
for the Galaxy obtained from our analysis in 
Figure 6.  Also, Table 1 gives the best-fitting
$2/n$ values obtained at different radii along with the total
gas-fraction at each radius. It shows that $2/n$ lies between 0 and 1 in general, or that $n$ lies between $\infty$ and 2. Thus, {\it our model vertical distribution is steeper than the one-component isothermal case ($n$=1) at all radii}, and lies between an exponential and a sech profile. This is in direct conformity with the results obtained by
Peletiers et al (1997) for a complete sample of external, edge-on galaxies.

Interestingly, the best-fitting steepness index $n$ is more directly dependent 
on the total absolute gas surface density than on the gas fraction (see Table 1).
This needs to be studied for other galaxy types and gas mass ranges in a future paper.

\subsection {Luminosity profile of a multi-component stellar disk}

So far we have treated the stars to consist of a single, isothermal component and 
our dynamical model  gives the steep vertical density profile for this single 
stellar component when coupled to the HI and H$_2$ gas 
components. We confirm the validity of this assumption next. 

Alternatively, a steepening of stellar vertical luminosity profile could also occur 
if a superposition of
stellar components with different velocity dispersions is considered, as has been 
suggested qualitatively (de Grijs et al. 1997). 
We check and refute this idea by considering a coupling of the
G-K-M dwarfs which contain the main mass fraction
in the stellar disk, and giants which contain a much smaller disk mass fraction
but dominate the disk luminosity. This latter feature is well-known, namely
that the luminosity of the disk is dominated by giants which may not determine 
its dynamics as cautioned by Mihalas \& Binney (1981).
The giants span a large range of velocities covering a factor of 2, and at
a first glance it may seem that that would result in a non-isothermal stellar 
luminosity profile, or a profile steeper than sech$^2$.
However, as we show next, the net dynamics in a disk consisting of coupled dwarfs and giants is
largely determined by the dwarfs because they constitute the main disk mass component
and the two have comparable dispersions.

We treat the giants to consist of two separate components or populations characterized by 
vertical dispersions of 14 km s$^{-1}$ and 28 km s$^{-1}$ respectively, with 23 \%
of the mass surface density of giants in the higher velocity component, as was shown
in a recent study based on the {\it {Hipparcos}} data by Holmberg \& Flynn 
(2004).    We then apply our coupled, three-component disk model to the 
G-K-M dwarfs (treated as a single component) and the above two giant 
components. The total surface densities of dwarfs and giants are taken 
to be 13.5 and 0.19 M$_{\odot} pc^{-2}$ respectively as observed, see Table 4-16 in Mihalas \& 
Binney (1981).

 The resulting net stellar density distribution from our model is 
fitted by the van der Kruit (1988) function to obtain the "n" parameter. 
Our dynamical model gives the $n$ value for dwarfs in this case to 
be 1.01, which is almost identical to a one-component sech$^2$ profile 
(which is given by n=1). For each of the two populations of giants, the best-fit $n$ 
value is also 1.01. Next, we add the density profiles for these two giant populations, 
and the best-fit value for $n$ in this case is found to be 1.07, which is still close 
to the isothermal case, although it is very slightly steeper than the profile for the 
individual components. The net luminosity profile, as set by the sum of the giant 
populations, will thus be nearly a sech$^2$ or close to an isothermal. This is a somewhat 
surprising result and physically this can be explained by the fact that the high-velocity 
giant population has a much lower surface density, hence in the coupled three-component 
system treated here, its density contribution within 150 pc (this is the $z$-range over which 
the fitting for $n$ is done) is much smaller than the other giant component. This is why despite 
the two giant populations having such different vertical velocity dispersions, the net density 
profiles of the giants is close to  that of a single isothermal. The $n$ for the combined giants 
is thus very similar to the $n$ for the main mass component namely the dwarfs.

    Note that the observed stellar luminosity profile is much steeper with
n=2 or higher, that is, it lies between a sech and an exponential (Section 3.2). 
Thus, a multi-component stellar disk with observed parameters for the 
dwarfs and the giants cannot explain the steep light distribution seen in galaxies. 
In contrast, the inclusion of gas in our model despite its small mass fraction
naturally results in the steep stellar profile as observed, as  shown in Section 3.2.

This calculation also confirms that our assumption of a single-component stellar disk 
is valid for the dynamical study, since the giants do not affect the dynamics of dwarfs.

Similarly, even in the coupled stars-gas case treated in our paper, 
the luminosity profile of the giants at low z can be taken to trace that of main stellar-mass 
component, namely the dwarfs.

\subsection {Model Surface Brightness vs. Height, $z$ }

The luminosity density of stars in a particular
colour-band is the directly observed quantity in astronomy rather
than mass density itself. But as the luminosity density in any
colour-band is taken to be a constant multiple of the stellar
density function, we conclude that the vertical luminosity distribution
follows the same functional form as the corresponding mass density
distribution, and hence is characterized by the same value of the
steepness index n.

We next obtain the surface brightness distribution of a
typical edge-on galaxy like our Milky Way as seen by an external
observer by integrating the luminosity density function along a
line of sight. If the luminosity density at a galactocentric
radius R, and vertical distance z above the mid-plane is given by
(see van der Kruit \& Searle 1981 a)

$$L(R,z)=L_0 \: sech^{2/n}(nz/2z_{e})  exp(-R/h) \eqno (9) $$
where 'h' is the disc scale-length, then the surface brightness is
given by:

$$ I(R,z)=I(0,0)(R/h) \: K_1(R/h) \: sech^{2/n}(nz/2z_e) \eqno(10) $$

\noindent where I(0,0)=2$L_0$h and $K_1$ is the
modified Bessel function. If I(R,z) is in units
of L$_{\odot}$ pc$^{-2}$, the surface brightness in units of
magnitude arcsec$^{-2}$ is (Binney \& Merrifield 1998):

$$
\mu(R,z)=-2.5 log_{10}(I/L_{\odot}pc^{-2})+M_\odot +21.572 \eqno (11) $$
\noindent Here $M_{\odot}$ denotes the absolute solar magnitude in the
chosen colour-band. For the K-band, $M_{\odot}$= 3.33 (Cox 2000),
 and the above formula reduces to

$$
\mu(R,z)=-2.5 log_{10}(I/L_{\odot}pc^{-2}) + 24.902 \eqno (12) $$
Our Galaxy is known to have an exponential disc (e.g., Mera et
al. 1998) with an exponential
 luminosity profile as given in equation (9), from this we obtain the
surface-brightness profiles in K-band at a given 
radius as follows. The stellar density value at each z (in units
of M$_{\odot}$pc$^{-3}$ ) at a given R obtained from our numerical
analysis is divided by M$_{\odot}/ $ L$_{\odot}$ 
of 1 to obtain the corresponding luminosity density in the K-band in units of
L$_{\odot}$pc$^{-3}$. This ratio of mass-to-luminosity is obtained using
 the K-band disk luminosity from Kent et al. (1991)
and the disk mass from Binney \& Tremaine (1987), and is in agreement with the study based on a large number of galaxies
(Bell \& de Jong 2001) which has shown the typical value of
this ratio to be $\sim 0.5$ with a scatter of a few around it.
Then using eqs. (10) and (11), we obtain the
surface-brightness distribution at a given radius. In Fig. 7, we
plot the model surface-brightness distribution in the K-band calculated at different
 radii, treating the Milky Way Galaxy as an edge-on system
as seen by an external observer.
These curves are strikingly similar to the observed profiles for
external edge-on galaxies (see Fig. 3   from Barteldrees \& Dettmar 1994;
and Fig. 5 from  de Grijs et al. 1997).

\section{Discussion}

\noindent 1. We have assumed the stellar disc to be isothermal for simplicity
as is the standard practice in the study of the
vertical structure of galactic disks (e.g., Spitzer 1942, Bahcall 1984). 
The stellar velocity dispersion increases due to secular
heating by tidal interactions and molecular clouds (e.g.,
Binney \& Tremaine 1987). Despite this the resulting stellar vertical velocity 
dispersion is shown to be constant with height (Jenkins \& Binney 1990, Jenkins 1992), 
hence the isothermal assumption made is robust.

\noindent 2. The model developed here is general, and since the gas velocity 
dispersion in external galaxies is universally much smaller than the stellar 
dispersion (see e.g., Lewis 1984, Wilson \& Scoville 1990), 
the analysis in this paper can be applied to other galaxies as well. We predict that the 
late-type galaxies being gas-rich will show  steeper stellar profiles (higher $n$ or lower $2/n$ values) than the early type galaxies, for the same range of radial range.
This could be checked with future observations.

\section{Conclusion}
It is well-established that the vertical distribution of 
stars in galactic disks is observed to be steeper than the sech$^2$  
profile expected from a one-component, isothermal
case, especially closer to the galactic mid-plane. 
In this paper, we provide a simple, physical origin for this:
we show that on taking account of the gas gravity,  
the resulting stellar vertical profile even under the isothermal assumption, 
is steeper than the sech$^2$ function. The resulting profile is calculated for our 
Galaxy using a three-component model of a galactic disc consisting of 
gravitationally coupled  stars and gas, and is shown to be steeper than sech$^2$ 
and lies between the sech profile and an 
exponential profile, in agreement with observations of galaxies. 
The higher the gas surface density, the steeper is the
resulting stellar vertical profile, asymptotically approaching the exponential    
curve. Due to the lower velocity of gas it is concentrated close to the mid-plane, this
crucial feature results in the steepening of the stellar profile.
We also show that 
the observed steep stellar profiles cannot be
explained by a superposition of several stellar components, as has sometimes 
been proposed qualitatively in the literature.

\noindent {\bf Acknowledgements:}
 We thank the anonymous referee for insightful comments which have improved 
our paper.


\clearpage

\begin{table}
\centering
  \begin{minipage}{140mm}
   \caption{Best-fitting $2/n$ values at different radii}
\begin{tabular}{llllll}
Radius & $\Sigma_{HI}$ & $\Sigma_{H_2}$ &  $\frac{\Sigma_{HI + H_2}}{\Sigma_s}$ &  n & 2/n  \\
(kpc)& M$_{\odot} pc^{-2}$  & M$_{\odot} pc^{-2}$ &&&\\
&&&&&\\

2 & 1.8 & 4.0 & 0.02  &3.33&0.6\\
3 & 3.8 &  4.9 & 0.04 & 5.4&0.37\\
4 & 4.6 & 13.1 & 0.10 &  7.32&0.27\\
4.5&  4.6 & 19.7 & 0.16 &9.12&0.22\\
5& 4.6 & 14.2& 0.14&7.91&0.25\\
6& 4.6 &  10.8& 0.16&6.89&0.29\\
7& 4.7  &4.9&  0.13  &4.86&0.41\\
8& 5.3 &3.6&0.17& 4.5&0.44\\
8.5& 5.6 &2.1&0.17&4.26&0.47\\
9& 5.6&1.4&0.18&4.22&0.47\\
10&5.6 &0.8&0.23&2.93&0.68\\
11&5.6 &0.6&0.30&2.88&0.69\\
12&5.6 &0.4&0.40&2.83&0.71\\
\end{tabular}
\end{minipage}
\end{table}

\clearpage

\begin{figure}
\vbox to0.6in{\rule{0pt}{2.6in}}
\epsscale{.8}
\plotone{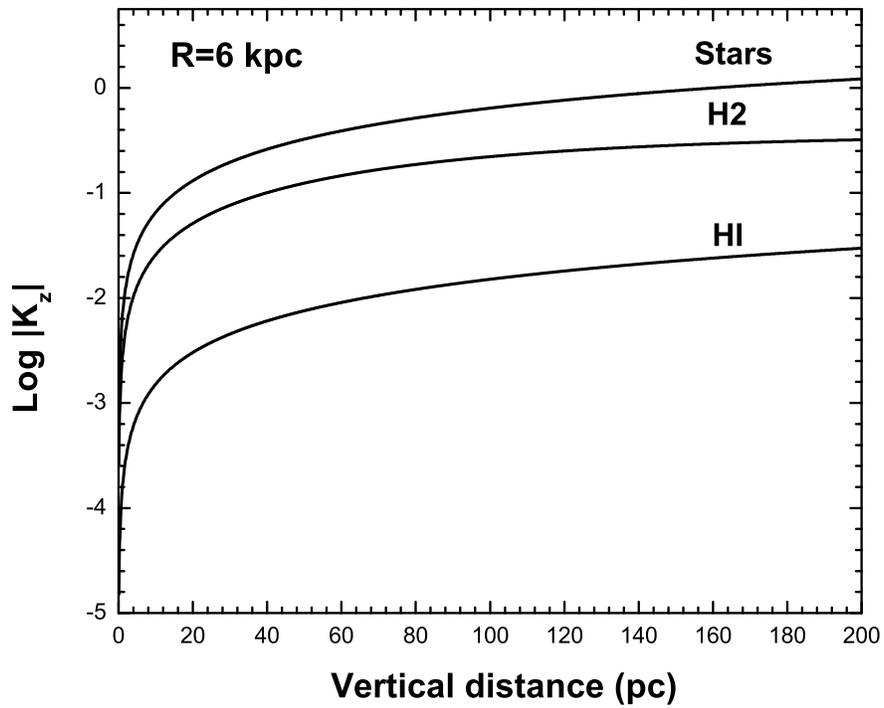}
\vskip 0.1in
\caption{Plot of force per unit mass $\vline K_z \vline$ 
for the star-alone case and the H$_2$-alone and HI-alone cases on a log-scale vs. z , calculated at R= 6 kpc.
The force due to the gas is a significant fraction $\sim 30 \% $ of the stars-alone case
for the z values close to the galactic mid-plane ($\vline z \vline \leq 150 $ pc). \label{fig1}}
\end{figure}

\clearpage

\begin{figure}
\vbox to0.6in{\rule{0pt}{2.6in}}
\epsscale{.8}
\plotone{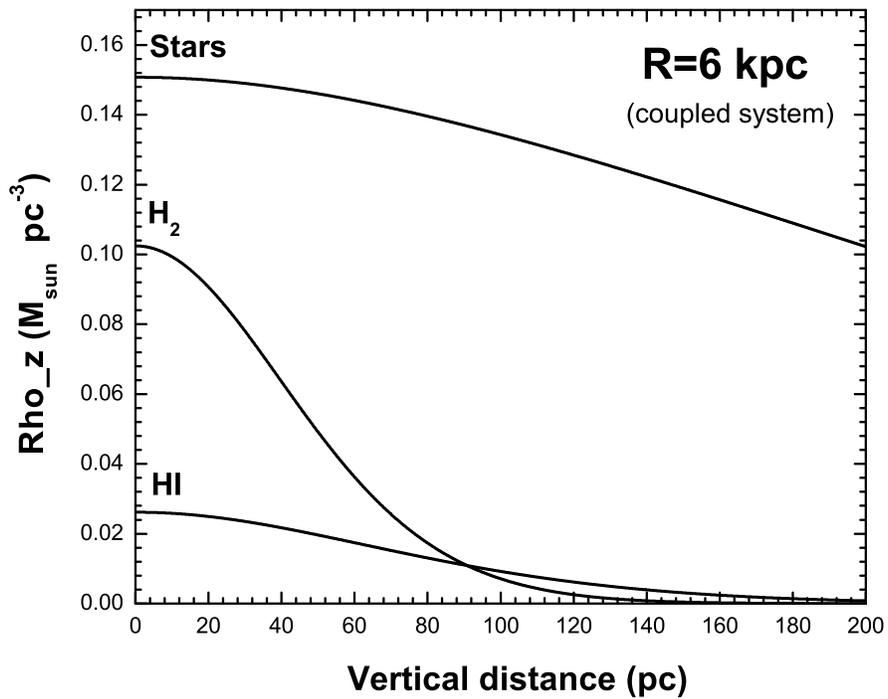}
\vskip 0.1in
\caption{The density distribution vs. z
for the gravitationally coupled three-component system obtained at R= 6 kpc, plotted 
 for the stars,  H$_2$, and HI gas components.
The gas density is a significant fraction of the stellar density close to mid-plane,
hence the force due to gas has a strong effect on the stellar distribution despite the 
fact that the gas surface density is a small fraction of the total disk surface density.
\label{fig2}}
\end{figure}

\clearpage

\begin{figure}
\vbox to0.6in{\rule{0pt}{2.6in}}
\epsscale{.8}
\plotone{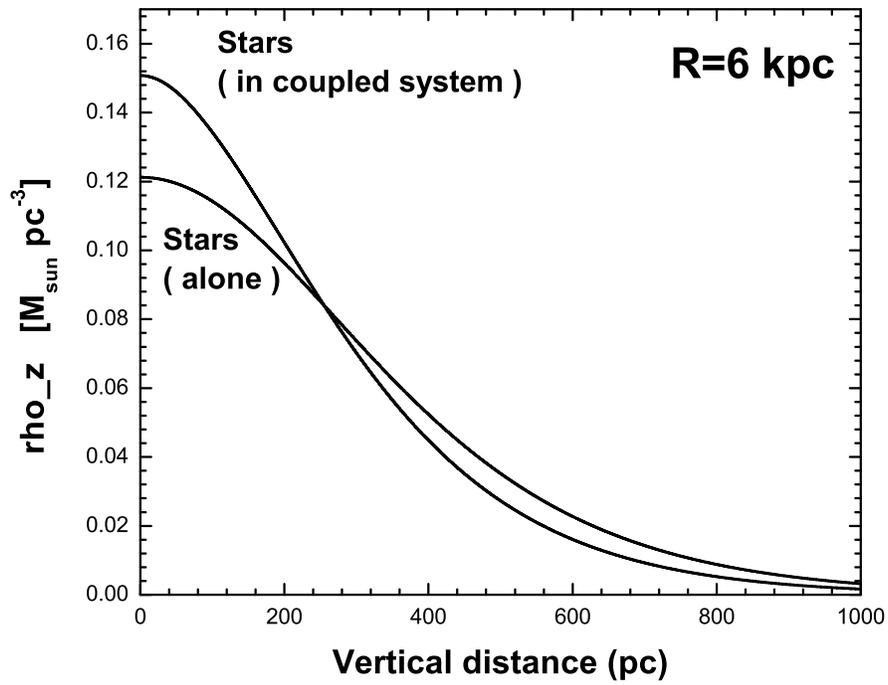}
\vskip 0.1in
\caption{The density distribution vs. z
for the star-alone case and 
 for the stars in a 
gravitationally coupled three-component 
system obtained at R = 6 kpc. 
Because of the gravitational force due to gas, the stellar distribution in the coupled case 
has a higher central value, the profile is steeper, and the
thickess (HWHM) is smaller.
\label{fig3}}
\end{figure}

\clearpage

\begin{figure}
\vbox to0.6in{\rule{0pt}{2.6in}}
\epsscale{.8}
\plotone{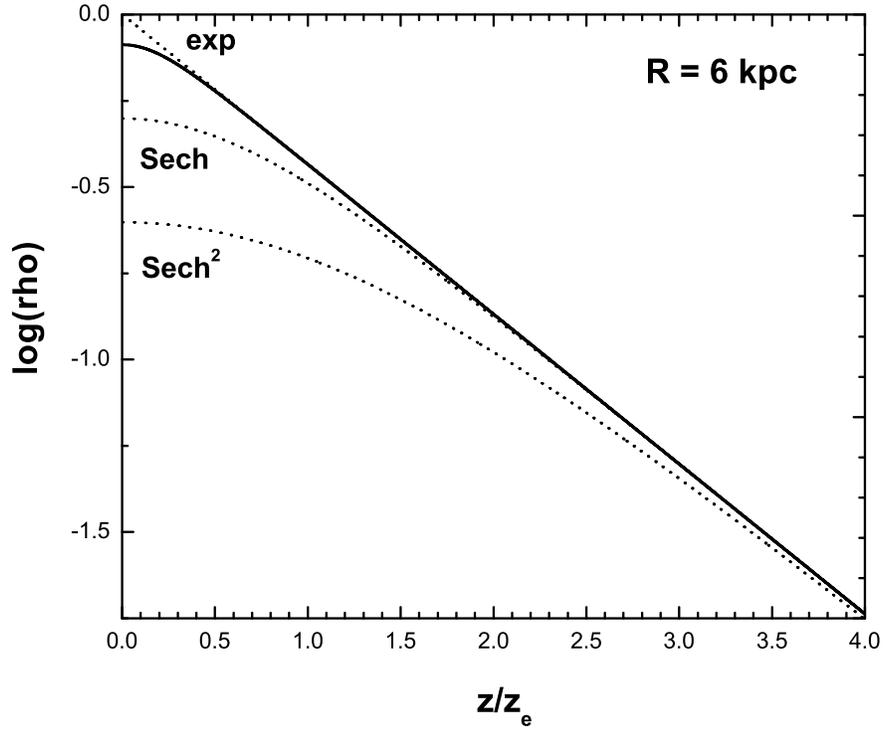}
\vskip 0.1in
\caption{Plot of vertical density distribution versus $z/z_e$, at R = 6 kpc. 
The model profile (solid line) is best-fitted by the n = 6.9 case from van der
 Kruit (1988), and is steeper than the stars-alone isothermal or the sech$^2$ case (n=1), and lies 
between the sech z (n=2) and an exponential (n=$\infty$) cases (shown as dotted lines). 
This region lies within the molecular ring of high
gas density , hence it
results in a steep vertical profile. \label{fig4}}
\end{figure}

\clearpage

\begin{figure}
\vbox to0.6in{\rule{0pt}{2.6in}}
\epsscale{.8}
\plotone{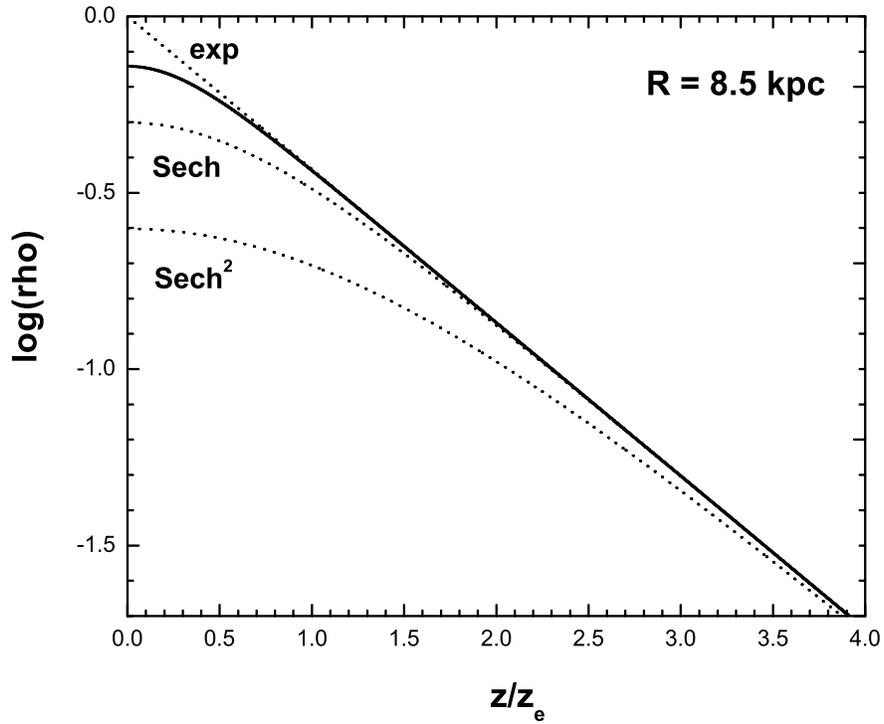}
\vskip 0.1in
\caption{Plot of vertical density distribution versus $z/z_0$, at R = 8.5 kpc. 
The vertical model profile is best-fitted by n=4.3 case from van der Kruit (1988), which is closer to the
sech case and is less sharp than at R = 6 kpc, but is still sharper than 
the stars-alone, isothermal case (sech$^2$ z). \label{fig5}}
\end{figure}

\clearpage

\begin{figure}
\vbox to0.6in{\rule{0pt}{2.6in}}
\epsscale{.8}
\plotone{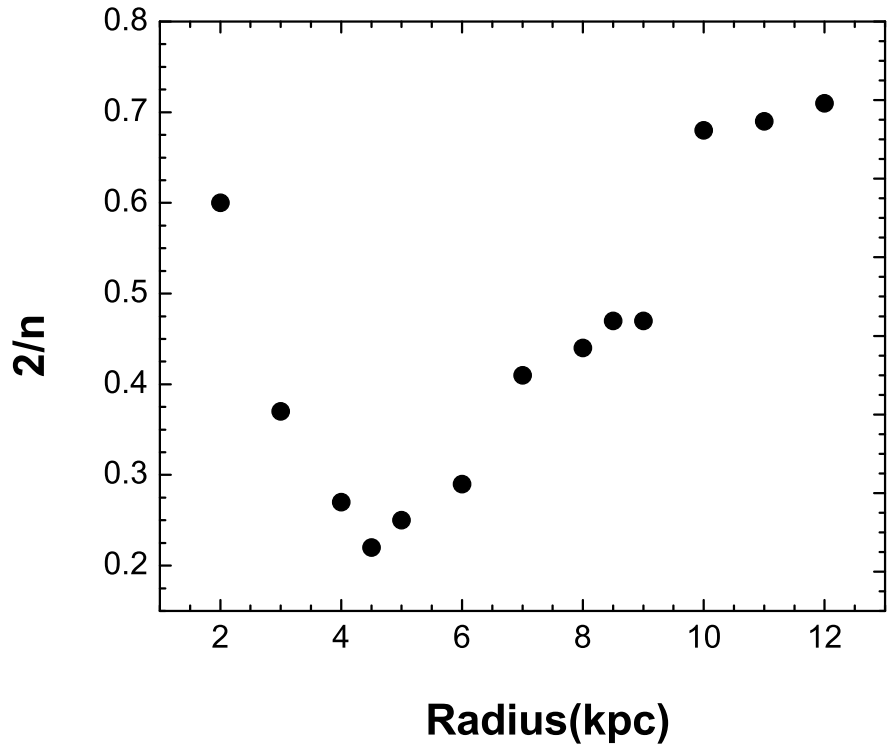}
\vskip 0.1in
\caption{A plot of the best-fitting $2/n$ values versus the radius between 
2-12 kpc for the Galaxy, where "n" is the van der Kruit parameter denoting the
steepness of the stellar vertical profile. \label{fig6}}
\end{figure}

\clearpage

\begin{figure}
\vbox to0.6in{\rule{0pt}{2.6in}}
\epsscale{.8}
\plotone{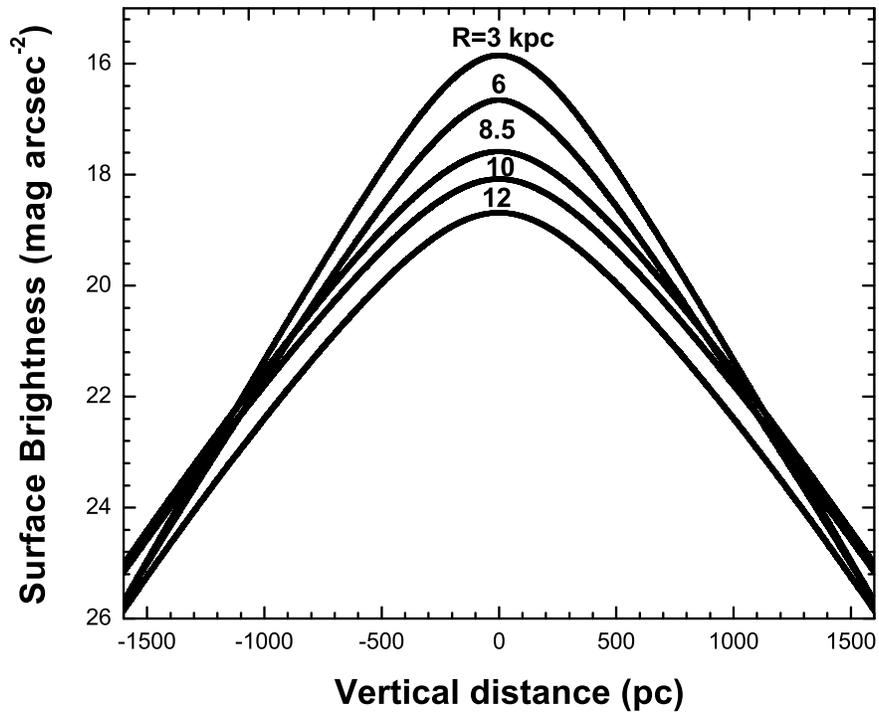}
\vskip 0.1in
\caption{A plot showing the model surface brightness curves in 
the K-band versus the distance z from the mid-plane, calculated at different 
galactocentric radii, treating the Galaxy as an edge-on system as seen by an 
external observer. \label{fig7}}
\end{figure}

\end{document}